\documentclass{sig-alternate-10pt}

\pdfoutput=1

\usepackage{microtype}

\usepackage[T1]{fontenc}
\usepackage[utf8]{inputenc}

\usepackage{graphicx}

\usepackage{url}

\usepackage{amsmath}
\usepackage{amssymb}
\usepackage{textcomp}

\usepackage{algpseudocode}
\algrenewcommand\algorithmicindent{1.0em}%
\usepackage{listings}

\usepackage{caption}
\usepackage[labelformat=simple]{subcaption}
\DeclareCaptionLabelSeparator{periodspace}{.\quad}
\usepackage{url}

\usepackage{cleveref}

\usepackage{stackengine}[2013-09-11]

\begin{document}

\title{ZeroSDN: A Message Bus for Flexible and Light-weight Network Control Distribution in SDN\titlenote{Technical Report TR-2016-06}}

\numberofauthors{1}

\author{
\alignauthor Frank D\"urr, Thomas Kohler, Jonas Grunert, Andre Kutzleb\\
\affaddr{University of Stuttgart}\\
\affaddr{Institute of Parallel and Distributed Systems (IPVS)}\\
\affaddr{Universit\"atsstra\ss{}e 38}\\
\affaddr{70569 Stuttgart}\\
\affaddr{Germany}\\
\texttt{frank.duerr@ipvs.uni-stuttgart.de}\\
\texttt{thomas.kohler@ipvs.uni-stuttgart.de}\\
\texttt{jonas.grunert@gmx.de}\\
\texttt{andre.kutzleb@googlemail.com}\\
}

\maketitle

\begin{abstract}
Recent years have seen an evolution of SDN control plane architectures, starting from simple monolithic controllers, over modular monolithic controllers, to distributed controllers. We observe, however, that today's distributed controllers still exhibit inflexibility with respect to the distribution of control logic. Therefore, we propose a novel architecture of a distributed SDN controller in this paper, providing maximum flexibility with respect to distribution.

Our architecture splits control logic into light-weight control modules, called \emph{controllets}, based on a \emph{micro-kernel} approach, reducing common controllet functionality to a bare minimum and factoring out all higher-level functionality. Light-weight controllets also allow for pushing control logic onto switches to minimize latency and communication overhead. Controllets are interconnected through a message bus supporting the publish/subscribe communication paradigm with specific extensions for content-based OpenFlow message filtering. Publish/subscribe allows for complete decoupling of controllets to further facilitate control plane distribution.

\end{abstract}

%

%
%
\section{Introduction}
Software-defined Networking (SDN) is based on the paradigm of \emph{logically centralized control} of network elements. Logical centralization is nothing more (but also nothing less!) than the concept of \emph{distribution transparency}, which is well-known from distributed systems. Basically, distribution transparency hides the complexity of a physically distributed system from the application by making distribution aspects ``transparent'', i.e., \emph{not}
visible to the application. Thus, the client can be implemented as if the system were centralized. In particular, network control applications implementing network control logic have a global view of the network, although network information such as topology information inherently has to be acquired through monitoring by distributed network elements (the switches). Moreover, the SDN controller itself might be (ideally) a distributed system with all its defining properties like replication transparency, fragmentation transparency, and without a single point of failure. For instance, topology information stored in a ``network information base'' might be replicated to and partitioned between many servers to ensure availability and scalability. 

\subsection{Evolution of SDN Controller Architectures}
Many SDN controllers have been implemented so far based on the concept of logically centralized control. Figure~\ref{fig:controller-archs} depicts the evolution of controller architectures. First SDN controllers were \emph{monolithic systems} basically implementing the controller as one process. The SDN controller connects through the southbound interface to the switches using, for instance, the popular OpenFlow protocol \cite{openflow-spec}, and the control applications interface with the SDN controller through a northbound interface, e.g., a Java API or REST interface. To increase fault-tolerance, the monolithic process implementing all control logic can also be fully replicated.

Very similar to the evolution of monolithic operating system kernels like the Linux kernel, this monolithic design was soon extended to a \emph{modular monolithic design} (Fig.~\ref{fig:sm_replication_modularized}), where control modules implementing certain control functions can be dynamically (un-)loaded into the controller process at runtime, e.g., using OSGi \cite{osgi}. One popular example showing that this design is still frequently used in practice is the OpenDaylight \cite{opendaylight} controller. However, similar to Linux still remaining a monolithic kernel also with kernel modules, we can also consider this modular controller architecture to be monolithic since it still relies on a central controller executing all modular control functions in one process. Again, the logically centralized controller can be physically distributed with each replica containing all control functions, i.e., replicas are identical clones.

Similar to the design of distributed operating systems, also SDN controller evolution continued to investigate \emph{distributed SDN controllers} (Fig.~\ref{fig:sm_partitioning}). Network control can be distributed along two dimensions. First, similar to the modular monolithic design, individual control functions can be factored out into control modules, which are now partitioned between different physical machines instead of fully replicating all control functions on all machines. Note, that this partitioning over control functions, as depicted, mandates multiple concurrent controller connections, which is not supported by OpenFlow. Secondly, control can be partitioned over the network topology, i.e., the scope of individual control modules can be restricted to certain switches. This possibly requires further concepts to coordinate instances with different scopes, e.g., through a controller hierarchy. 

\subsection{Towards a Distributed Micro-Kernel Architecture for SDN Controllers}
Observing that distributed SDN controllers already exist today, can we conclude that SDN controller evolution has reached its end? We argue that this is not the case, for the following reasons.

First of all, direct communication between network elements to implement \emph{fully distributed network control} protocols (without switch-external control functions) is not anticipated. With current SDN, switches talk to the external monolithic network controller or distributed external control modules, but not to other switches. This reflects the clean-slate paradigm shift from distributed network control to logically centralized control, where switches are just fast forwarders and all ``intelligence'' is outsourced to external machines. Obviously, this outsourcing comes at a cost like increased round-trip times (slower reaction), increased control network load, or difficult implementation of robust logically centralized control relying on additional machines that can fail.  
Therefore, we argue that a truly flexible SDN architecture would allow for the full spectrum of distribution, from fully centralized to fully distributed control. In other words: we must enable the possibility to bring control back onto the switch.

Secondly, with the current concept we observe that controllers tend to be quite heavy-weight (which might also be a practical reason why control is removed from switches). A good example might be the prominent OpenDaylight controller. In order to just receive packet-in events, OpenDaylight requires a full-fledged OSGi environment and total code size of $\approx 280\,\textrm{MByte}$. We argue that it should be possible to identify a minimal feature set that every control module can implement, basically to communicate with switches and other distributed control modules. Anything else should be factored out into the implementation of the control function. In other words, we advocate a light-weight \emph{micro-kernel} approach for SDN controllers instead of a heavy-weight monolithic controller architecture.

Thirdly, we observe that switches and controllers are still tightly coupled, which hinders the free distribution of control logic. For instance, OpenFlow requires a TCP connection to one master controller and possibly one slave controller. Since TCP is inherently based on connections to certain machines, spawning new control applications at other machines or migrating them between machines is cumbersome and potentially disruptive \cite{krishnamurthy_pratyaastha:_2014,basta_towards_2015}. We argue that switches must be \emph{decoupled} from the SDN controller. This can be achieved by using state-of-the-art communication middleware approaches as already successfully used in other domains for the communication between services \cite{chappell2004enterprise}. As a side effect, choosing a suitable communication middleware also allows for implementing control logic in virtually any language and to support event-driven as well as request/response types of interaction (in contrast to, for instance, the RESTful paradigm, which is language independent but not supporting eventing). 

The main contribution of this paper is a novel architecture for a distributed SDN controller fulfilling all of the above requirements: (1) high flexibility with respect to distribution of control logic covering the full spectrum from logically centralized to fully distributed control; (2) micro-kernel controller architecture for distributed light-weight controller modules (so-called controllets); (3) push-down of controllets implementing control logic onto switches; (4) decoupling of controllets through a message bus supporting content-based filtering of so-called data plane events. A first implementation of the proposed concepts is publicly available under the name \emph{ZeroSDN} on GitHub \cite{github-zsdn}.

The rest of the paper is structured as follows. In Section~\ref{sec:architecture}, we describe the architecture of our distributed SDN controller together with an overview of the basic concepts. In Section~\ref{sec:sdn_message_bus}, we describe the message bus concept in more detail. In Section~\ref{sec:control_plane_distribution}, we discuss how this concept enables highest flexibility in terms of control plane distribution, before we describe some details of our preliminary implementation of the concepts in Section~\ref{sec:priliminary-work}. Based on the presented concepts for distributed network control, we outline a roadmap towards a highly scalable holistic system control plane in Section~\ref{sec:extensions}, before we discuss related work in Section~\ref{sec:related-work}, and conclude the paper in Section~\ref{sec:summary}.

\section{Architecture}
\label{sec:architecture}
\begin{figure*}[t]%
\begin{subfigure}[t]{.32\linewidth}
   \centering
	 \includegraphics[height=4.25cm]{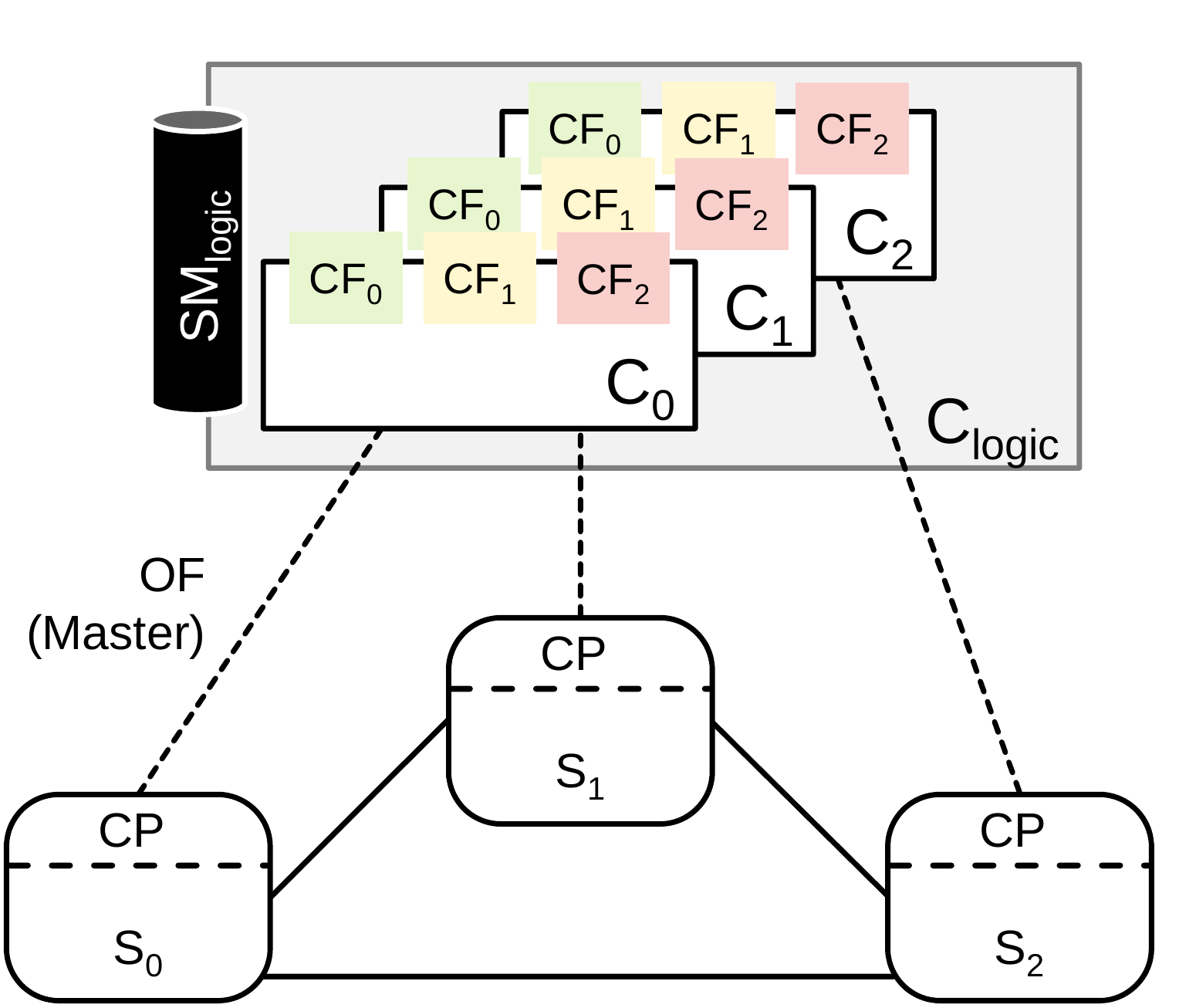}
   \caption{\small Modular, monolithic, replicated;\\\textit{C: controller instance, CF: control function, SM: state module}}
   \label{fig:sm_replication_modularized}
\end{subfigure}
\hspace*{2mm}
\begin{subfigure}[t]{.32\linewidth}
   \centering
	 \includegraphics[height=4.25cm]{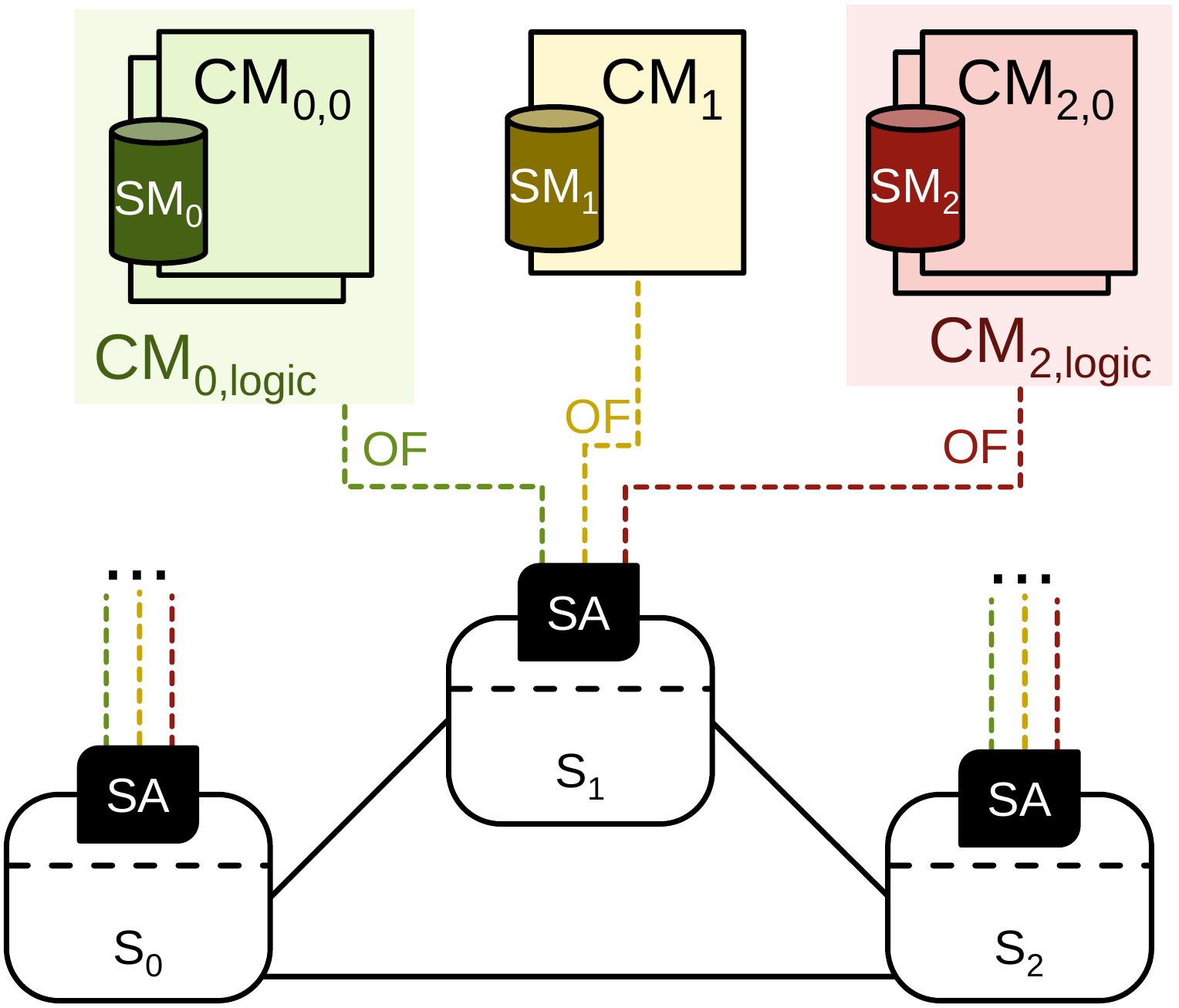}%
   \caption{\small Modular, distributed, partitioned;\\\textit{CM: control module, SA: switch adapter}}
   \label{fig:sm_partitioning} 
\end{subfigure}
\hspace*{2mm}
\begin{subfigure}[t]{.32\linewidth}
   \centering
	 \includegraphics[height=4.25cm]{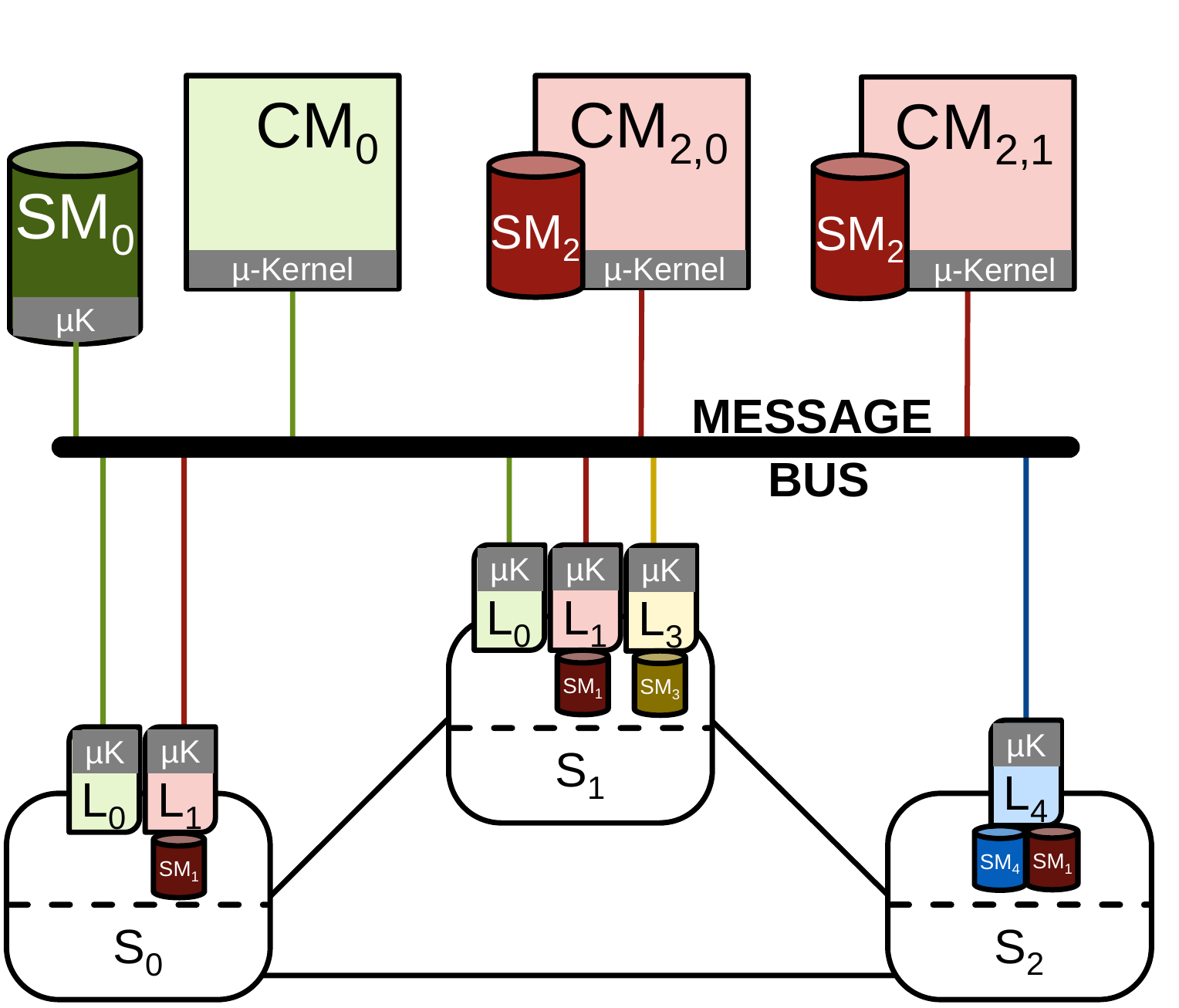}%
	 \caption{\small \textmu-kernel architecture with fully distributed local \textit{(L)} \& external controllets \textit{(CM)}, interconnected by a message bus.}
   \label{fig:sm_fully_distributed} 
\end{subfigure}
\vspace*{2mm}
%
\caption{Evolution of distribution in SDN controller architectures. Rightmost: Our envisioned fully distributed architecture.}
\label{fig:controller-archs}%
\end{figure*}%
We start by introducing the basic architecture of our distributed SDN controller (cf. Fig.~\ref{fig:sm_fully_distributed}) including an overview of the basic functions and concepts.

Our approach is based on what we call a \emph{micro-kernel architecture} for SDN controllers. The basic idea is to split network control logic into light-weight control modules, whose instances we call \emph{controllets}. In contrast to a monolithic controller, controllets do not require a heavy-weight execution environment such as an OSGi framework (e.g., in case of OpenDaylight), and they are also not executed in the same process. Instead we execute each controllet in a separate process and enable each controllet to communicate with other controllets or switches through messages. The micro-kernel just provides basic functions for messaging including publish/subscribe message routing and parsing (in particular of OpenFlow messages), and registration and discovery of controllets and switches. Any other functionality like network topology management, routing, etc. is implemented by the controllets' ``business'' logic. One advantage of having a slim functionality for the SDN micro-kernel is that we can port the micro-kernel with little effort to different languages enabling us to basically use any language for the implementation of controllets. Moreover, the light-weight nature of controllets also enables us to execute controllets directly on switches, typically featuring limited computing resources, to push-down control logic onto switches. This decreases communication latency and overhead.

Communication is based on a \emph{message bus} to decouple controllets from switches and other controllets. Each controllet and switch can send messages through the bus to other controllets or switches either using the request/response or the publish/subscribe (pub/sub) paradigm. In particular pub/sub communication decouples controllets from switches. Switches publish events like packet-in events, and controllets subscribe to events, and vice versa. The sender does not actually know which other components are receiving the events. It is the task of the message bus to route messages to the correct receiver(s), i.e., subscriber(s). To this end, we also include specific content-based filters to subscribe to OpenFlow messages based on the match fields. Decoupling controllets and switches allows for flexible distribution including migration of controllets, and dynamic spawning or exchanging of controllets at runtime.  

Overall, this architecture allows for maximum flexibility. Next, we elaborate on the technical details and further features enabled by this approach.

\section{The SDN Message Bus: Decoupling Controllets through Events}
\label{sec:sdn_message_bus}
Our architecture utilizes \emph{event-based communication} to decouple the producers of events from the consumers of events in time (asynchronous communication) and space (distribution of logic between nodes including switches and hosts). In the domain of SDN, we in particular consider so-called \emph{data plane events}, stemming from packets or state changes of data plane elements (switches and end systems). Data plane events include the common examples like the addition or removal of network elements, link status updates, or the arrival or departure of certain packets at switches. However, we do not restrict ourselves to such \emph{basic data plane events}, but also explicitly consider \emph{complex data plane events} involving, for instance, several packets and timing conditions. For instance, a complex event could be triggered by a certain sequence of packets, or the non-arrival, i.e., absence, of a certain packet over a defined period of time. Typically, switches only fire basic events, which are then forwarded to subscribing controllets, which in turn evaluate complex event conditions to fire complex data plane events.

The SDN message bus is responsible for routing event notifications to their subscribers. Since data plane events often include matches on packet header fields, we argue that the message bus should support \emph{content-based filtering} of events \cite{eugster_many_2003}. Therefore, event conditions include matches on header field tuples or any other meta-data. This paradigm can also emulate standard client/server communication, multicast, or topics \cite{eugster_many_2003}, using filters on receivers, groups, topics, etc.
 
Recent SDN research has shown that consistency in an inherently distributed system of switches and controllers might require certain semantics on the delivery of messages (``exactly once'' delivery) \cite{katta_ravana:_2015}. The message bus paradigm is well-suited to transparently implement this strong semantic together with more light-weight semantics like ``at most once'' delivery for less critical tasks. The necessary code for the publishers and subscribers is part of the micro-kernels included by every controllet.     

Events are heavily used by distributed control, as described next.

\section{Truly Flexible Control Plane Distribution}
\label{sec:control_plane_distribution}
Common SDN architectures treat switches as ``dumb'' network elements that can do fast forwarding, but have to be configured by an ``intelligent'' remote controller implementing the network control logic. On the one hand, this reduces the functionality of switches to a bare minimum. On the other hand, outsourcing control from the switch increases latency due to round-trip times and network overhead due to remote communication. In this section, we will show how light-weight controllets can bring back control onto the switch and still benefit from the logically centralized paradigm of SDN.

\subsection{Fully Distributed Control}
\label{sec:fully_distributed_control}
SDN has abandoned fully decentralized network control based on a distributed control plane implemented solely by switches in favor for logically centralized control, at least if we consider a ``clean slate'' SDN network without hybrid switches also implementing distributed control protocols alongside logically centralized network control. We do not want to argue for or against logically centralized control or fully distributed control. However, we observe that the strict notion of separating data plane elements (switches) and the logically centralized control plane (external network controller) imposed by OpenFlow limits the full potential and flexibility of the SDN paradigm. For instance, ``legacy'' distributed control protocols, such as distance vector or link state routing protocols, have proven to be fault-tolerant and scalable. As investigated by \cite{katta_ravana:_2015}, vigorous efforts have to be undertaken to provide the same fault-tolerance with a logically centralized SDN network. Therefore, we stress the fact that maintaining a global view and applying logically centralized control algorithms comes at a cost, and the advantages of logically centralized control should be well traded-off against its disadvantages. Consequently, we argue that truly flexible network control also includes the option for full distribution of network control, to let the network operator decide what paradigm fits his needs best.

Moreover, new developments in networking hardware and virtualization make it feasible to push control logic onto switches. \textit{Whitebox Switches} with open, Linux-based network operating systems (NOS) and increasingly powerful CPUs\footnote{For instance, one prominent recent Whitebox Networking Switch features an x86 Atom Rangley CPU with 4 cores at $2.4\,\textrm{GHz}$, $8\,\textrm{GB}$ RAM (DDR3), and $8\,\textrm{GB}$ FLASH memory.}, similar to server CPUs, enable the execution of custom control logic directly on the switch. Light-weight virtualization technologies like containers or unikernels facilitate the deployment and isolation of logic on switches with such open NOS.

Therefore, and in-line with recent research \cite{schmid_exploiting_2013,bianchi_openstate:_2014,bifulco_improving_2016,cascone_traffic_2015}, our architecture supports pushing light-weight controllets directly onto the switch, as illustrated in \ref{fig:sm_fully_distributed}. These switch-local controllets can then execute a fully distributed network control protocol, e.g., for decentralized routing without involving a logically centralized controller anymore. Like any controllet, also switch-local controllets communicate through the message bus using events---thus, we can implement distributed network control alongside logically centralized network control, or implement anything in-between as shown in Sec.~\ref{sec:central_coordination}. 

For the case where a switch is not powerful enough to execute a controllet, we also support executing the controller on a host and running just a slim switch adapter on the switch to connect to the external controllet.
 
\subsection{Centrally Coordinated Distributed Control}   
\label{sec:central_coordination}
Fully distributed control is one extreme from the spectrum of control plane distribution. The other end is fully (logically) centralized control based on a global view. Seeing benefits in both extremes, we asked ourselves how to get the best of both worlds? We name this distributed control scheme \emph{centrally coordinated distributed control}. 

The basic idea of this scheme is to calculate multiple alternative plans how individual switches should react in certain situations centrally on a switch-external controllet based on a global view. These plans are then distributed to switch-local controllets together with the event conditions that should trigger alternative plans. Switch-local controllets subscribe to the given events and trigger adaptations based on the pre-calculated plans without further switch-controller round-trip. Events are delivered directly to switches avoiding triangular routing via the controller. 

Defining a set of distributed plans and conditions that guarantee a consistent distributed behavior is challenging, and a general concept for this problem is currently under investigation by us. Here, we just want to show that our proposed SDN architecture also supports such advanced distributed control concepts. 

%
%
\section{Implementation}
\label{sec:priliminary-work}
To demonstrate the major concepts, we have implemented a proof-of-concept prototype called \emph{ZeroSDN} \cite{github-zsdn}. The core of ZeroSDN is implemented in C{}\verb!++!, although control modules (controllets) can be implemented in practically any language as discussed below. ZeroSDN employs the production-grade communication middleware ZeroMQ \cite{zeromq}. The implementation of ZeroSDN is split into two parts: 

\begin{enumerate}
\item a modular execution framework
\item the distributed SDN controller application, implemented on top of the execution framework by a set of controllets.
\end{enumerate}

The \emph{execution framework} offers automatic controllet discovery along with dependency and life-cycle management of controllets. Controllets communicate through the ZeroMQ messaging middleware, which provides a high-performance, low-latency message bus supporting both, publish/subscribe communication for efficient event filtering as well as direct request/reply communication.

The \emph{distributed SDN controller application} includes controllets implementing the core functionality of the SDN controller, such as topology management and simple forwarding. To this end, ZeroSDN supports OpenFlow version 1.0 and 1.3. Controllets can be implemented in any language supported by the ZeroMQ messaging middleware, so it is practically language-independent. For efficient message de-/serialization, we use Protocol Buffers \cite{protobuf}.

In the following, we focus on one crucial aspect of our implementation: how to map data plane events onto hierarchical event topics enabling controllets to subscribe to relevant data plane events delivered through ZeroMQ\footnote{Additional explanations and examples are given in the Github Wiki \cite{github-zsdn}.}? We illustrate the usage of pub/sub messages below with the example of the \texttt{SwitchAdapter} (SA) controllet.

Each controllet defines two sets of \textit{topics}:

\begin{enumerate}
\item Set \texttt{TO} defines which message types (topics) a controllet is able to process, i.e., which data plane events it wants to receive from the message bus. This set is mapped to corresponding subscriptions for filtering event messages.
\item Set \texttt{FROM} defines the topics published by the controllet, i.e., events sent to the message bus. Other controllets can subscribe to these advertised topics.
\end{enumerate}

Topic definition is strictly hierarchical. The first hierarchy layer defines the type of declaration (\texttt{TO} or \texttt{FROM}). The second layer comprises the identity of the controllet. All upper layers contain the structure of controllet-type specific content. Attributes are encoded as a bit-sequence, with a specific length associated to each hierarchy layer, at a specific location within the topic hierarchy. Wildcard matching (``\texttt{?}'') is supported.

As an example, we described the event topics of the \emph{SwitchAdapter (SA)} controllet in more detail. The SA wraps an OpenFlow-enabled switch to connect the switch to the ZeroSDN message bus. The SA is either executed locally on a whitebox switch or remotely on a server to connect blackbox switches to the message bus. A switch connects to an SA instance through OpenFlow as it would normally connect to an SDN controller. From the perspective of the switch, its SA is its controller. From the perspective of any other controllet, an SA instance represents an OpenFlow switch. 

Figure~\ref{lst:sa_topics} shows the declaration of the topics of the SA. The SA subscribes to several OpenFlow messages (\texttt{TO}) that need to be transmitted from the control applications (other controllets) via the SA to the switch, such as packet-out or flow-mod messages. The SA will match an incoming pub/sub message from the message bus with these topics and forwards matching messages to the switch. 

The SA publishes (\texttt{FROM}) several events to the message bus. From the perspective of the SA, these are events created from OpenFlow messages, such as packet-in messages, received from the switch that are then sent by the SA via the message bus to the control application (other controllets).

\begin{figure}
\begin{lstlisting}
TO=0x01
  SWITCH_ADAPTER=0x0000
    SWITCH_INSTANCE=0x????????????????
      OPENFLOW=0x00
       PACKET_OUT=0x0D
       FLOW_MOD=0x0E
       ...
FROM=0x02
  SWITCH_ADAPTER=0x0000
    OPENFLOW=0x00
      PACKET_IN=0x0A
        LB_GROUP=0x??  default=0x00
	  IPv4=0x0800
	    TCP=0x06
	    UDP=0x11
	    ...
	  ARP=0x0806
	    ...
      PORT_STATUS=0x0C
            ...
\end{lstlisting}
\caption{Excerpt of the \texttt{SwitchAdapter} topic-hierarchy.}
\label{lst:sa_topics}
\end{figure}

Note that hierarchy layers are not restricted to directly reflect OF-matching fields. Artificial hierarchy layers may be freely introduced between any layers. For instance, to enable load balancing of \texttt{PACKET\_IN} messages, the SA artificially discriminates \texttt{PACKET\_IN}s by introducing an additional 1-Byte topic hierarchy layer (\texttt{LB\_GROUP}), to disseminate such events in a round-robin fashion to the set of \texttt{LB\_GROUP} members. Controllets participating in load balancing subscribe to a specific \texttt{LB\_GROUP}, whereas controllets that want to receive all \texttt{PACKET\_IN}s apply a wildcard subscription on the \texttt{LB\_GROUP} layer.

Preliminary evaluations show that our approach scales linearly with the number of controllet replicas. With 4 replicas running on dedicated nodes (Xeon $4 \times 2.4\,\textrm{GHz}$, $24\,\textrm{GB}$ RAM), we achieve a throughput of $\approx 500\textrm{k}$ messages per second (Cbench \cite{cbench} experiment, 16 switches).

\section{Roadmap towards a Highly Scalable Holistic System Control Plane}
\label{sec:extensions}
We also would like to discuss how to further improve and leverage the presented concept of event-driven distributed network control.

Our distributed SDN controller is based on content-based filtering of events, in particular, the filtering of data plane events based on header field matching. In larger networks, event notifications might arrive at a high rate, which makes content-based message filtering by the message bus challenging. In our prototype, we used a high-performance topic-based messaging system (ZeroMQ) as a workaround by mapping match fields onto a topic hierarchy. However, such a topic mapping also comes with inherent problems. In particular, attributes have to be specified according to the order given by the topic hierarchy. Irrelevant hierarchy levels can be wildcarded, however, efficient wildcard topic matching at high event rates is hard to implement in software.

To solve this problem, we observe that content-based data plane event filtering is conceptually similar to matching header fields during packet forwarding by switches in the data plane. Hardware switches achieve line-rate forwarding performance in the data plane through special hardware (TCAM) supporting also wildcard matching efficiently. So one interesting question is, could we utilize similar hardware for implementing an \emph{SDN message bus appliance} supporting content-based event filtering and routing to subscribing controllets? Similar appliances have already been used in other domains like service-oriented architectures, where XML appliances speed-up XML processing (transformation, filtering, and routing of XML traffic) \cite{pasetto_design_2012}.

As a second extension, we can further leverage the publish/subscribe paradigm to build a \emph{holistic distributed system controller} not limited to controlling the network elements but to include virtual network functions, end systems (including virtual machines), applications (e.g., client and server processes on the application layer), etc. In other words, we can extend the network control plane to a \emph{holistic system control plane} implemented by a set of distributed controllets, which communicate indirectly through events including not only data plane events but any event relevant for controlling the holistic system. As a simple example, consider the migration of a virtual machine (VM), which might also require the migration of virtual network functions like firewalls, and the adaptation of routes for chaining services. Using event-based communication, we can trigger actions to implement an event-triggered workflow defining the sequence of actions necessary to migrate the VM. For instance, as soon as the VM has been suspended by a VM controllet, an event could be fired that triggers the migration of network functions, which then trigger the adaptation of routes in the network through further events. This way, complex \emph{system management workflows} can be implemented in a decentralized fashion. 

%
%
\section{Related Work}
\label{sec:related-work}
Many early approaches, including Onix \cite{koponen_onix:_2010}, propose to externalize state storage, which incurs additional latency for lookups. In Onix, switches and controller instances are tightly coupled. While Onix limits the shared view onto network state information, HyperFlow \cite{tootoonchian_hyperflow:_2010}, as our approach, holistically propagates all kinds of data plane events. HyperFlow also facilitates pub/sub to propagate events, event classification is however limited to three topics, whereas our approach leverages content-based filtering (mapped to a topic hierarchy in our preliminary implementation) to allow for fine-grained subscriptions. Furthermore, HyperFlow exclusively relies on passive synchronization of the locally cached network wide view, while our approach offers maximum flexibility allowing both, local caches for fast access as well as access to highly consistent centralized storage.

DevoFlow \cite{curtis_devoflow:_2011} is the first SDN approach to allow for local decision making on the switch, however mandating changes of the switching ASIC. Kandoo \cite{hassas_yeganeh_kandoo:_2012} proposes a two-layered controller hierarchy with a root controller maintaining network-wide state, and local controllers possibly running directly on switching hardware, only handling local events where no global knowledge is required. While this scheme allows for offloading of simple local logic, local controllers do not hold any state data, neither do they interact with each other at all. Our approach is not limited to such a strict hierarchical scheme and does not rely on a root controller instance, thus offering superior flexibility.

While these approaches exhibit a static switch-controller assignment, ElastiCon \cite{dixit_elasticon:_2014} allows for a dynamic switch to controller instance mapping. By periodic monitoring of controller load, the number of instances and the mapping is adapted for effective load-balancing. Since switches are still tightly coupled to an instance, the authors introduce a switch migration protocol. A similar problem is addressed in \cite{krishnamurthy_pratyaastha:_2014,basta_towards_2015}. The decoupling of switch and controller offered by our approach eliminates the need for complex and costly migration mechanisms entirely.

More recent approaches improve on failure tolerance in control distribution. Beehive \cite{yeganeh2016beehive} models control applications as centralized asynchronous message handlers featuring and thus focusing on application partitioning, exclusive handling of messages among a set of controllers, as well as consistent replication of control state information. Logical message propagation is dictated by map-functions that determine to which set of applications a specific message is to be sent to. Message passing is not addressed in detail. Furthermore, each Beehive controller instance contains all application logic in contrast to our highly modular approach.
Another work, Ravana \cite{katta_ravana:_2015}, focuses on controller fault-tolerance. Ravana subsumes event dissemination from switches, their processing by a controller, and the resulting execution of controller commands at the switches in a transaction and guarantees that control messages are processed transactionally with exactly-once semantics. Message propagation and actual distribution schemes are not addressed there.

Fibbing \cite{vissicchio_central_2015} exerts centralized control over routers that implement a legacy, non-SDN, control plane running fully decentralized routing algorithms, such as OSPF and IS-IS. The forwarding behavior of routers, i.e., their forwarding information base, is manipulated as to achieve desired network behavior by faking input messages to the distributed routing algorithms. Although being congruent in the notion of centralized control, unlike in our approach, Fibbing's control is solely indirect and thus inherently limited.

%
%
\section{Summary}
\label{sec:summary}
In this paper we presented a novel architecture for a highly flexible distributed SDN controller based on a message bus for communication and a micro-kernel design. Network control logic is split into control modules, called controllets, that can be freely distributed. Controllets communicate through the message bus and are decoupled from switches and other controllets using the publish/subscribe paradigm. The micro-kernel design only requires controllets to implement a small set of functions to connect to the message bus and participate in publish/subscribe communication. Consequently, controllets are extremely light-weight and can also be executed directly on Whitebox Switches to enable fully distributed network control even without external SDN controller---a new level of flexibility in control plane distribution that so far is not possible with standard SDN controllers.

We also identified future research directions like hardware-assisted processing of data plane events to further increase the scalability of content-based event filtering in the control plane, or the event-driven execution of system management workflows in a holistic system control plane embracing the network control plane.

\bibliographystyle{abbrv}
\bibliography{bibliography}
\label{last-page}

\end{document}